\documentclass[10pt,journal,compsoc]{IEEEtran}
\pdfoutput=1
\usepackage{amsfonts}
\usepackage[ngerman,english]{babel}
\usepackage{mathtools}
\usepackage{mathabx}
\usepackage{amsmath}

\usepackage{graphicx}
\usepackage{color}
\usepackage{tikz}
\usepackage{pgfplots}
\usepackage{pgf-umlsd}
\usepackage{ifthen}
\usepackage[]{fp}
\usepackage{booktabs}
\usepackage{multirow}

\usepackage{float}
\usepackage{ragged2e}

\usetikzlibrary{matrix,patterns,spy,fit,calc}
\usepgfplotslibrary{groupplots}
\pgfplotsset{compat=newest}

\FPset{totalOffset}{0}

\ifCLASSOPTIONcompsoc
  \usepackage[nocompress]{cite}
\else
  \usepackage{cite}
\fi
\ifCLASSINFOpdf
\else
\fi
\usepackage{amsmath}
\begin{document}

%
\title{Deep Iteration Assisted by Multi-level Obey-pixel Network Discriminator (DIAMOND) for Medical Image Recovery}
%
%
%
%

\author{Moran~Xu,~
        Dianlin~Hu,~
		Weifei~Wu$^*$,
        and~Weiwen~Wu
\IEEEcompsocitemizethanks{\IEEEcompsocthanksitem M. R. Xu and W. F. Wu are with the People's Hospital of China Three Gorges University, Yichang, 443000, China and they are also with the First People's Hospital of Yichang, Yichang, 443000, China. D. L. Hu are with the Laboratory of Imaging Science and Technology, School of Computer Science and Engineering, Southeast University, Nanjing, 211189, China. W. W. Wu is with the department of radiology diagnosis in the Univeristy of Hong Kong, 999077, SAR, China.($^*$ refers to the corresponding author)\protect\\
E-mail: xumoran34@gmail.com, dianlinhu@gmail.com, \\ 
wuweifei236@sina.com, weiwenwu12@gmail.com
\iffalse \IEEEcompsocthanksitem J. Doe and J. Doe are with Anonymous University.\fi}
\iffalse \thanks{Manuscript received April 19, 2005; revised August 26, 2015.}\fi}

\IEEEtitleabstractindextext{%
\begin{abstract}
\justifying Image restoration is a typical ill-posed problem, and it contains various tasks. In the medical imaging field, an ill-posed image interrupts diagnosis and even following image processing. Both traditional iterative and up-to-date deep networks have attracted much attention and obtained a significant improvement in reconstructing satisfying images. This study combines their advantages into one unified mathematical model and proposes a general image restoration strategy to deal with such problems. This strategy consists of two modules. First, a novel generative adversarial net(GAN) with WGAN-GP training is built to recover image structures and subtle details. Then, a deep iteration module promotes image quality with a combination of pre-trained deep networks and compressed sensing algorithms by ADMM optimization. (D)eep (I)teration module suppresses image artifacts and further recovers subtle image details, (A)ssisted by (M)ulti-level (O)bey-pixel feature extraction networks (D)iscriminator to recover general structures. Therefore, the proposed strategy is named DIAMOND.
\end{abstract}

}

\maketitle

\IEEEdisplaynontitleabstractindextext

%
\IEEEpeerreviewmaketitle

\IEEEraisesectionheading{\section{Introduction}\label{sec:introduction}}

%
%
%
%
\IEEEPARstart{I}{mage} recovery is a significant part of inverse problems. Specifically, when an original image is polluted by noise, the recovery task aims to remove noise and preserve fine details. When the original image is blurred because of motions, the task mainly focuses on recovering a sharp image from the deblurring one. When the original image does not satisfy resoution demand, the recovery task is transferred to resolution enhancement, etc. In this study, we focus on a common approach to solve image recovery tasks. Specifically, image denoising and image super-resolution tasks are discussed.

\subsection{Image Super Resolution}

There are two types of image super-resolution reconstruction technology. One is to synthesize a high-resolution image from multiple low-resolution images, and the other is to obtain a high-resolution image from a single low-resolution image. In this column, we focus on Single Image Super-Resolution Reconstruction (SISR). SISR methods can be divided into three categories: interpolation-based methods, reconstruction-based methods, and learning-based methods. Interpolation-based methods are simple to implement and have been widely used, but these linear models limit their ability to recover high-frequency details. Sparse representation\cite{yang2008image} based technologies enhance the ability of linear models by using prior knowledge. This type of technology assumes that any natural image can be sparsely represented by a dictionary's elements. This dictionary can form a database and learn the mapping from low-resolution images to high-resolution images from the database. However, such methods are computationally complex and require many computing resources\cite{wu2020dictionary}\cite{hu2019sister}. Based on CNN (Convolutional Neural Network) model, SRCNN\cite{dong2015image} first introduced CNN into SISR. It only used a three-layer network and achieved advanced results. Subsequently, various models based on deep learning entered the field of SISR, roughly divided into the following two significant directions. One is to pursue the recovery of details, using PSNR, SSIM, and other evaluation standard algorithms, among which the SRCNN model is the representative. Another is a series of algorithms represented by SRGAN\cite{ledig2017photo} and ESRGAN\cite{wang2018esrgan}, which aims to reduce the perceptual loss without paying attention to details and looking at the big picture. The two algorithms in different directions have different application fields. In medical imaging, the details and features of the image is helpful for making a precise diagnosis instead of pursuing the image's overall clarity. Therefore, in this work, we will dig into the algorithms that pursue detail restoration and their medical field applications. Algorithms pursuing detail restoration are also sorted into three categories.

1. Pre-sampling super-resolution: this algorithm uses traditional interpolation as a preprocessing to obtain coarse higher-resolution images and then refines them using deep neural networks\cite{dong2015image}\cite{kim2016accurate}\cite{tai2017memnet}\cite{tai2017image}\cite{kim2016deeply}\cite{shocher2018zero}. 

2. Post-sampling super-resolution: Most computation is performed in low-dimension,  the predefined upsampling is replaced with end-to-end learnable layers integrated at the end of the models\cite{ledig2017photo}\cite{lim2017enhanced}\cite{tong2017image}\cite{han2018image}.

3. Progressive upsampling super-resolution: the networks are based on a cascade of CNNs and progressively reconstruct higher-resolution images. At each stage, the images are upsampled to higher resolution and refined by CNNs\cite{lai2017deep}\cite{wang2018fully}\cite{ma2017learning}.

The SRCNN model\cite{dong2015image} is a pioneering work of introducing deep learning into SISR and using bicubic interpolation as the preprocessing process. Subsequently, the VDSR model\cite{kim2016accurate} introduced the residual structure to SISR. Instead of directly learning the mapping from low-resolution images to high-resolution images, VDSR learns the residuals of the two images. Residual learning structure not only accelerates convergence speed of model training but also introduces deeper network structure into SISR so that the model has a wider receptive field. The DRCN\cite{kim2016deeply} model introduces the recursive structure into the SISR and divides the model into three areas. One is the Embedding network, the other is the Inference network, and the third is the Reconstruction network. The highlight of this model lies in the intermediate Inference network and loss function. Inference network shares a convolution parameter, there are D layers, the output of each layer is pooled together, and then two losses are defined. The first type, local loss, is the difference between each layer's output value and HR image. The second type computes the difference between the weighted average output of all layers and the HR image, combining these two losses to form the overall loss. The FSRCNN model\cite{dong2016accelerating} uses deconvolution to replace the interpolation in the SRCNN model and directly learns the mapping from low-resolution images to high-resolution images to achieve end-to-end training. The core concept of ESPCN\cite{shi2016real} is sub-pixel convolutional layer (also called "pixel shuffle"). The input of the network is the original low-resolution image. After passing through three convolutional layers, a featured image with $r^2$ channels equal to the input image size is obtained. Then rearrange each channel pixel of the feature image into an $r\times r$ area, corresponding to a sub-block of size $r^2$ in the high-resolution image, so that the feature size is $H\times W\times r^2$ Images are rearranged into ${rH}\times {rW}\times 1$ high-resolution images. The sub-pixel convolutional layer proposed by the ESPCN model is widely used in subsequent studies. Compared with the deconvolutional layer of the FSRCNN model, it can learn the nonlinearity of low-resolution to high-resolution images. The SRDenseNet model\cite{tong2017image} introduces DenseNet into the SISR field. DenseNet inputs the features of each layer in the dense block to all subsequent layers so that the features of all layers are concatenated, instead of directly performing tensor summation like in ResNet\cite{he2016deep}. This architecture brings the advantages of reducing the problem of gradient disappearance, strengthening feature propagation, supporting feature reuse, and reducing the weight parameters to the entire network. Hu et al. \cite{hu2019runet}deploys Resblocks into U-Net \cite{ronneberger2015u} architectures to enhance video resolution and suppresses blurring. Recently, the Generative Adversarial Nets (GANs) \cite{goodfellow2014generative} attract much attention to resolution enhancement because of its advantages in better promoting finer details, sharp edges, and removing inaccurate artifacts. SRGAN\cite{dong2015image} first introduced GAN into the field of super-resolution. Compared with traditional GAN study, SRGAN inputs low-resolution image instead of noise samples. Moreover, SRGAN defines a Content loss, which is a weighted sum of MSE loss and perceptual loss, to replace a simple MSE loss in generator training. Together with adversarial loss, the author named the whole generator loss as 'Perceptual Loss'. By making full use of a pre-trained network like VGG19 in ImageNet, Perceptual loss can complement texture information in high-resolution outputs. Based on SRGAN, Wang et al. \cite{wang2018esrgan} proposed an enhanced version of that structure called ESRGAN, which outperforms SRGAN in many SR competitions. Comparingly, ESRGAN accesses four promotions: First, ESRGAN replaced Residual Block with Dense Blocks and removed all Batch Normalization in training. Batch Normalization is similar to a kind of contrast stretching for images. After any image passing through Batch Norm, its color distribution will be normalized. In other words, it destroys the original contrast information of the image, which is not suitable for pixel-level image generation tasks like this study. Compared with ResBlocks, Dense Blocks intends to converge to a globally optimal solution, especially without BN constraint. Second, the discriminator part of the loss function is modified by subtraction between real data loss and generated data loss. Third, generator loss was calculated using the characteristic map before ReLU activation. Finally, Using the network interpolation method to settle the contradiction between the objective evaluation index and subjective visual effect.

\subsection{Image Denoising}
The importance of image denoising in low-level vision can be revealed in many aspects:

1. Noise corruption is inevitable during image acquisition and processing, and it will heavily degrade image quality and add interference to high-level vision tasks.

2. In medical imaging, even subtle noise may misguide diagnosis.

3. In step-progressive inference via splitting variables, many image restoration tasks can be addressed by embedding an intermediate denoising step, further expanding its application fields.

Image denoising technology has become a research hotspot for tens of years. For example, using non-local similarity\cite{buades2005non} to optimize the sparse method can improve denoising performance. Dictionary learning\cite{mairal2008supervised} helps remove noise quickly. The prior knowledge\cite{levin2011natural} \cite{babacan2012bayesian} \cite{romano2017little} restores the details of the potentially clean image by smoothing the noise image. More competitive denoising methods including BM3D\cite{dabov2007image}, WNNM\cite{gu2014weighted}, NLR-MRF\cite{sun2011learning} and TNRD\cite{chen2015learning} can be used. Although most of these methods achieved good performance in image denoising, they have the following drawbacks:

1.The testing phase involves complex optimization methods.

2.Numerous manually set parameters.

3.Denoising models are fixed to certain denoising tasks.

Deep learning technology with strong self-learning ability can address these shortcomings. The application of deep learning technology in image denoising includes deep learning technology of additive white noisy(AWNI) image denoising, deep learning technology of real noise image denoising, deep learning technology of blind denoising, and deep learning technology of composite noise image denoising. DnCNN\cite{zhang2017beyond} proposes to use convolution to learn from end-to-end residuals and the perspective of functional regression, using convolutional neural networks to separate noise from noisy images and achieve denoising results that are significantly better than other methods. Since then, a series of improvements based on the network structure have been proposed. Residual Encoder-Decoder Network(REDNet)\cite{jiang2018rednet} uses a deep convolutional encoding-decoding framework based on symmetric skip connections, so that in the reverse process, information can be directly transferred from the top layer to the bottom layer; Memory Network(MemNet)\cite{tai2017memnet} further proposes a long-term memory model for image denoising; Multi-level Wavelet CNN(MWCNN)\cite{liu2018multi} proposes a multi-level wavelet CNN framework, which is beneficial to restore image details by combining discrete wavelet transform with a convolutional network. The above methods usually require separate training models for different noise levels, which not only lack flexibility but also cannot be applied to real noise images with more complex degradation processes. CBDNet\cite{guo2019toward} is a blind denoising method that combines noise estimation and a non-blind denoising model. By relying on signal-dependent noise and the influence of camera image signal processing on noise, synthetic noise and real noise images are used for network training, which can achieve sound denoising effects and generalization capabilities on real noise images. A new trend to combine the traditional mathematical model with deep learning priors has become a hotspot. For example, Regularization by Denoising (RED)\cite{romano2017little}, and its more efficient variant \cite{sun2020block} attempt to incorporate deep learning priors into denoising models and achieve relatively good performances. In image denoising tasks, convolutional neural networks have achieved great success. However, most of the existing models are based on noise-clear image pairs for supervised learning. In some specific applications, such as CT, MRI, due to the difficulty of obtaining explicit images, the methods based on unsupervised learning show a wide range of application prospects. However, the existing convolutional neural network denoising methods based on unsupervised learning are still in the trial stage, and the training speed and recovery performance need improvement. Therefore, it is of great significance to explore self-supervised and unsupervised learning methods for real noisy images.

In this study, we work on a strategy to deal with general image recovery tasks, including image denoising and image super-resolution. In imaging scenes like camera, CT, and MRI, we face a complex image recovery task. That is to say, to reconstruct a satisfying image, we need to deploy step-by-step techniques. For example, due to the X-ray transmitter's low-dose photon, the CT image will be polluted by severe noise. In the meantime, the image resolution may need to be enhanced to satisfy diagnosis demand because of the CT machine's resolution limit. Our research focuses on a general post-processing strategy to deal with such composite image recovery tasks. The main contributions are threefold: First, we combine deep learning module with deep iteration module to reconstruct different kinds of image recovery tasks by one strategy. Second, we propose a novel GAN network for the deep learning module to recover image details and lost information. Third, the previous network, together with a compressed sensing technique, is deployed further to promote image quality for the deep iteration module. The proposed method is proved to be effective in both super-resolution and denoising tasks.

The rest of the paper is organized as follows: In section II, we briefly review related mathematical theories and then establish the deep network module and deep iteration module of our proposed method. In section III, both image super-resolution and image denoising experiments on various datasets are performed, and then we compare several indexes to evaluate the proposed method qualitatively and quantitatively. In section IV, we discuss some related issues and conclude.

\section{Proposed Method}
The proposed method is useful in various image reconstruction tasks, including but not limited to, resolution enhancement and image denoising. This method consists of two modules. First, a generative adversarial network with WGAN-GP training is built to recover general image structures; Second, a post-processing strategy, named Iteration Refinement (IR), deploys a compressed sensing method and a pre-trained network to recover details and suppress artifacts iteratively. During training the proposed network, for denoising tasks, a set of input label images are first polluted by Gaussian white noise; for image super-resolution tasks, label images are downsampled by a factor of two in both $x$ and $y$ directions and then upsampled by a factor of 2 in both directions using Bicubic interpolation. For training the network, every low-quality image and its corresponding label image are used. In testing, given a low-quality input image, the trained network predicts the high-quality image. Then the predicted image is fed into an iterative module to promote image quality again iteratively.

\begin{figure*}
	\centering
	\includegraphics[width = 1.0\textwidth]{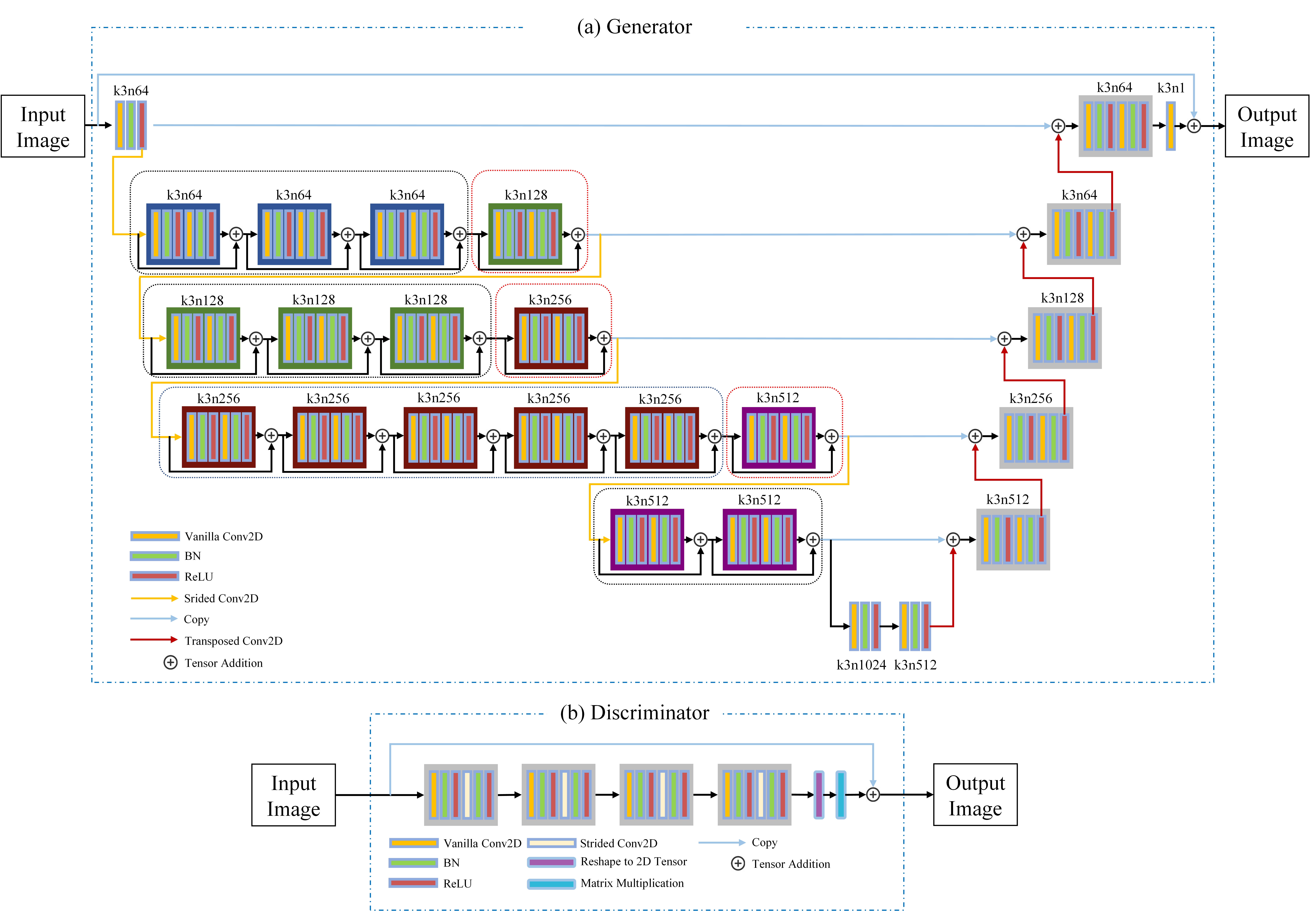}
	\caption{Architecture of (a) generator and (b) discriminator network. Where $kz_1nz_2$ represents a convolutional layer with kernel size of $z_1\times z_1$ and $z_2$ feature maps. The residual blocks in black dotted boxes is used in image super-resolution tasks, and should be replaced by two sets of convolution-batch norm-ReLU layers in denoising tasks. Similarly, residual blocks in red dotted boxes are suitable for super-resolution tasks, and should be replaced by two sets of convolution-batch norm-ReLU layers in denoising tasks.}
	\label{Figure:1}
\end{figure*}

\subsection{Network Architecture Module}
The proposed DIAMOND architecture consists of a generator subnetwork and a discriminator subnetwork and performs WGAN-GP training. For the generator, we base our network architecture on RUNet\cite{hu2019runet}, which is used initially to enhance image resolution in video sequences. To better apply RUNet into image recovery tasks, we use strided convolution in the contracting path to strengthen its multi-level feature extraction function. Apart from that, to reduce training parameters and speed up WGAN-GP training, we use pixel summation instead of original concatenation in the expanding path. For discriminator, we set up a multi-level feature extraction network with a one-dimension tensor output, suggested by WGAN-GP, to calculate and minimize Wasserstein Distance.

\subsubsection{Deep Generator Architecture}
Following Arjovsky et al.\cite{arjovsky2017wasserstein} and Gulrajani et al.\cite{gulrajani2017improved}, we define a discriminator network $D_\omega$ which we optimize in an alternating manner along with $G_\theta$ to solve the adversarial min-max problem:
\begin{equation}
\begin{aligned}\min_{\theta} \max_{\omega}\mathbb{E}_ { \textbf{I}\sim \mathbb{P}_g}[D_\omega(\textbf{I})]- \mathbb{E}_ {\textbf{I}\sim \mathbb{P}_r}[D_\omega(\textbf{I})]+{} \\
 \Lambda\mathbb{E}_ {\textbf{I}\sim \mathbb{P}_{\hat{\textbf I}}}[(\left \| \nabla_{ \textbf{I}}D_\omega(\textbf{I}) \right \|_2-1])^2]\end{aligned}
\end{equation} 
Suppose $\textbf{I}^L$ is the input low-quality image, $\textbf{I}^H$ is the high-quality label image. Where $\hat{\textbf I}=G_\theta(\textbf{I}^L)$, $\epsilon \sim Uniform[0,1]$, $\hat{\textbf I} =\epsilon \textbf{I}^H+(1-\epsilon)\hat{\textbf I}$, $\Lambda$ is the coefficient for gradient panelty term.

Unlike traditional GAN by Goodfellow et al.\cite{goodfellow2014generative}, WGAN-GP training stables the training process by removing logarithm in loss functions, discarding sigmoid activation in discriminators, and adding a gradient penalty term in loss functions. With this approach, our generator can learn to create highly similar solutions to real label images and thus difficult to classify by $D$. Meanwhile, the network is much easier to converge.

Our generator network consists of several residual blocks, strided convolutions, and tensor operations, as shown in Fig. 1(a). We use the residual training method to optimize the training process, which means the proposed generator learns the residue image between label images and low-quality images. Unlike conventional UNet architecture\cite{ronneberger2015u}, the contracting path (left path) shown in Figure 1(a) consists of a sequence of blocks, each followed by a tensor addition operation to feed forward the same block input to the subsequent block, so-called residual block\cite{lim2017enhanced}. This architecture allows the network to transfer shallow features directly to deep layers. The image features can be better preserved using multiple residual blocks in every step of the contracting path. To efficiently upscale the low-resolution image, the transposed convolution layers are used for the expanding path, the right path shown in Fig. 1(a). The number of residual blocks deployed in every step will be further discussed in 3.3.2. The residual blocks in black dotted boxes are used in image super-resolution tasks and replaced by two groups of convolution-BN-ReLU layers in denoising tasks. Similarly, residual blocks in red dotted boxes are suitable for super-resolution tasks and should be replaced by two groups of convolution-BN-ReLU layers in denoising tasks.

For the contracting path, the input residual image passes through one set of convolution-BN-ReLU layer to produce 64 feature maps, then its size is contracted by half.

Moreover, our proposed generator modifies classic RUNet from two aspects. First, instead of deploying pooling layers, we use convolutional layers with stride 2 and 1/2 for in-network downsampling and upsampling, which will enlarge the reception fields. Specifically, we utilize $k (=4)$ downsampling and upsampling steps in the modified RUNet, leading to $k+1$ spatial scales of feature maps. Second, we adopt a simple pixel-wise summation operation to combine the feature maps from the encoder and decoder subnetworks instead of concatenation utilized in UNet. We empirically find that element-wise summation effectively reduces the network parameters and can lead to comparable reconstruction results.

\subsubsection{Discriminator and WGAN-GP Training}
To achieve better perceptual performance, we use perceptual loss function\cite{johnson2016perceptual} during all training tasks, as shown in the following section. However, the perceptual loss has a severe shortcoming to introduce annular or rectangular artifacts in reconstruction images. According to our research, the proposed discriminator structure performs well in suppressing such artifacts and attain more delicate features. A well-trained discriminator indicates 'distance' from the generated image to the real image by minimizing discriminant loss (Wasserstein loss in our study). By alternately training a generator and discriminator, annular artifacts can be effectively suppressed during this process.

To discriminate real images from generated image samples, we train a discriminator network. The architecture is shown in Figure 1(b). We follow the guidelines from Radford et al.\cite{radford2015unsupervised} and use LeakyReLU activation $(\sigma=0.2)$ to avoid 'Dead Neurons'. Unlike the original method of using max-pooling to reduce image sizes, we applied strided convolution throughout the network to enlarge reception fields. The discriminator network is trained to solve the maximization problem in Equation 1. It contains eight convolutional layers with an increasing number of $3\times3$ filter kernels, increasing by a factor of 2 from 32 to 256 kernels. Strided convolutions are used to reduce image resolution and increase channels each time the number of features is doubled. The resulting 256 feature maps are followed by one dense layer to obtain a one-dimensional tensor for WGAN-GP training. By deploying a discriminator network, we can suppress annular artifacts introduced by perceptual loss function; By removing sigmoid function at the output layer, we follow WGAN-GP training demands, which means the training process can achieve better global convergence.

According to \cite{arjovsky2017wasserstein} and \cite{gulrajani2017improved}, WGAN-GP training is adaptable to various GAN training procedures. By removing sigmoid activation in the discriminator's output layer, discarding all logarithms in generator and discriminator losses, and adding a gradient penalty term to stabilize gradient descent, WGAN-GP training introduces Wasserstein Distance instead of Jensen-Shannon divergence in loss functions to prevent gradient vanishing problems. When applying WGAN-GP training, the exponential decay rate of first-moment estimation($\beta _1$) and second-moment estimation($\beta _2$) in discriminator's adam optimizer are empirically set as 0.5 and 0.9.

\subsubsection{Perceptual Loss Function}
The definition of our perceptual loss function $l$ is critical for our generator network's performance. While $l$ is commonly based on the MSE\cite{lim2017enhanced}, we consider the perceptual loss functions\cite{johnson2016perceptual} which map the predicted image $ \hat{\textbf I}$ and the target high-quality image $\textbf{I}^H$ into a feature space and measure the distance between the two mapped images in the feature space. We formulate the perceptual loss as the weighted sum of a content loss $(l_C)$ and an adversarial loss component$(l_{Gen})$ as :
\begin{equation}
l=l_C+\lambda l_{Gen}
\end{equation}
where $\lambda$ is a hyper-parameter. In the following we describe our choices for the content loss $(l_C)$ and adversarial loss $(l_{Gen})$.

\paragraph{Content Loss}
The pixel-wise MSE loss is calculated as:
\begin{equation}
l_{MSE}=\frac{1}{WHC} \sum^C_{z=1} \sum^W_{x=1}\sum^H_{y=1}(\textbf{I}_{x,y,z}^{H}-G_\theta(\textbf{I}^L)_{x,y,z})^2
\end{equation}
where $\textbf{I}^L$ is the input low-quality image. Above is the most widely used optimization target for image reconstruction tasks. However, while achieving exceptionally high PSNR value, reconstructions often lack sharp edges and fine details. In other words, the high-frequency contents of the image are not preserved, resulting in unsatisfying solutions with overly smooth textures.

To solve this problem, we rely on the ideas of Johnson\cite{johnson2016perceptual} et al. and use a loss function to measure perceptual similarity. We use a pre-trained VGG-16 network proposed by Simonyan and Zisserman \cite{simonyan2014very}. Let $\Phi = \{\phi_j, j=1,2,...,N_p\}$ denote a loss network that extracts features from a given input image and consists of $N_p$ convolutional layers, where $\phi_j(\textbf{I})$ denotes a feature map of size $C_j\times H_j \times W_j$ obtained at the $j^{th}$ convolutional layer for the input image $\textbf{I}^L$, and $N_p=5$ in this paper. Given a predicted image $\hat{\textbf I}$ and a target image $\textbf{I}^{H}$, the feature distance $\ell^j$ at the $j^{th}$ layer can be computed as follows:
\begin{equation}
\ell^j=\frac{1}{W_jH_jC_j}(\phi_j(\hat{\textbf I})-\phi_j(\textbf{I}^{H}))^2
\end{equation}
So, the content loss can be written as:
\begin{equation}
\l_{C}=\sum_{j=1}^{N_p}\ell^j
\end{equation}

\paragraph{Adversarial Loss}
As we mentioned before in 2.1.1, merely using perceptual loss will introduce annular or rectangular artifacts in reconstruction images. To solve this problem, we combine the perceptual loss with an adversarial loss to further suppress artifacts. By trying to fool the discriminator, this network encourages generated images to approach real images gradually. We absorb the idea from Arjovsky et al.\cite{arjovsky2017wasserstein} and Gulrajani er al \cite{gulrajani2017improved} to deploy WGAN-GP training. Specifically, the generator loss $l_{Gen}$ is defined as:
\begin{equation}
l_{Gen} = \sum _{n=1}^{N}-D_\omega(G_{\theta}(\textbf{I}^L))
\end{equation}
Here, $-D_\omega(G_{\theta}(\textbf{I}^L))$ is the loss function of generator loss in WGAN-GP training which indicates the probability that the reconstructed image $G_{\theta}(\textbf{I}^L)$ is a real image. Correspondingly, we use $D_\omega(G_{\theta}(\textbf{I}^L))-D_\omega(\textbf{I}^H)+\lambda(\left \| \nabla_{\hat{\textbf I}}D_{\omega}(\hat{\textbf I}) \right \|_2-1)^2$, where $\hat{\textbf I}=\epsilon \textbf{I}^H + (1-\epsilon)G_{\theta}(\textbf{I}^L)$ as the discriminator loss in WGAN-GP training.

\subsection{Deep Iteration Module}
For a given low-quality image $\textbf I^L$, and $\textbf I^{(k)}$ is recovered image at $k$th iteration using the deep iteration module, where $k\in[1, K]$ is the index of iteration, and $K$ is the total number of iterations. $H$ represents the operation kernel, which is blurred kernel and Gauss kernel in this study. $\psi(\textbf I)$ is a trained deep reconstruction network, which transfers a poor image quality to a good recovered image. $f(\cdot)$ is a regularization prior penalized on the recovered image. The goal of deep iteration module is to search the solution satisfying measurement data within the near domain of current iteration. In general, the optimization model based on current image is formulated as follows:
\begin{equation}
\begin{aligned}
\left \{\textbf I^{(k+1)}, \textbf y^{(k+1)}\right \}=argmin_{\left \{ \textbf I, \textbf y \right \}}(\frac{1}{2}\left \| \textbf y -(\textbf I^{(L)}-\textbf {HI}^{(k)}) \right \| _F^2 +{} \\ \frac{\mu}{2}\left \| \textbf I - \textbf I^{(k)}-\psi(\textbf y)\right \|_F^2+\xi f(\textbf I))
\end{aligned}
\end{equation}
where $\mu\textgreater0$ and $\xi\textgreater0$ are weighting parameters to balance the component of deep learning and regularization term. The first term on the right enforces data fidelity in the measurement domain. The second term emphasizes the recovered images need to satisfy the requirement of deep learning prior. The third term based on $f(\textbf I)$ is a general regularizer by considering the general priors. The mathematical model of Eq. (7) enables a superior image reconstruction based on a combination of a deep image prior and a regularization prior. 

Because the model of Eq. (7) contains the optimization of neural network, i.e., $\psi(\textbf y)$, which is complex and we can replacing $\psi(\textbf y)$ with $\textbf g$ and then Eq. (7) is converted to be the following form:
\begin{equation}
\begin{aligned}
\left \{\textbf I^{(k+1)}, \textbf g^{(k+1)}, \textbf y^{(k+1)}\right \}=
&argmin_{\left \{ \textbf I, \textbf g, \textbf y \right \}}(\frac{1}{2}\left \| \textbf y -(\textbf I^{(L)}-\textbf {HI}^{(k)}) \right \| _F^2 \\ 
&+\frac{\mu}{2}\left \| \textbf I - \textbf I^{(k)}-\textbf g\right \|_F^2+\xi f(\textbf I)), \\
&s.t. \textbf g=\psi(\textbf y)
\end{aligned}
\end{equation}
The mathematical model of Eq. (8) is a constraint optimization problem, which can be convert into the following unconstrained problem:
\begin{equation}
\begin{aligned}
\left \{\textbf I^{(k+1)}, \textbf g^{(k+1)}\right \}=
&argmin_{\left \{ \textbf I, \textbf g, \textbf y \right \}}(\frac{1}{2}\left \| \textbf y -(\textbf I^{(L)}-\textbf {HI}^{(k)}) \right \| _F^2 \\ 
&+\frac{\mu}{2}\left \| \textbf I - \textbf I^{(k)}-\textbf g\right \|_F^2+\xi f(\textbf I)), \\
&+\frac{\upsilon}{2}\left \| \textbf g - \psi(\textbf y)\right \|_F^2)
\end{aligned}
\end{equation}
where there are three variables to be optimized. By using the alternating optimization strategy, it can be divided into three sub-problem: the sub-problem of solving $\textbf y$, the sub-problem of solving $\textbf g$, the sub-problem $\textbf I$, which can be respectively written as follows:
\begin{equation}
\begin{aligned}
\textbf y^{(k+1)}=argmin_{\textbf y}(\frac{1}{2}\left \| \textbf y -(\textbf I^{(L)}-\textbf {HI}^{(k)}) \right \| _F^2 +{} \\\frac {\upsilon}{2}\left \| \textbf g^{(k)} - \psi(\textbf y)\right \|_F^2, 
\end{aligned}
\end{equation}
\begin{equation}
\begin{aligned}
\textbf g^{(k+1)}=argmin_{\textbf g}\frac{\mu}{2}\left \| \textbf I - \textbf I^{(k)}-\textbf g\right \|_F^2 +\frac {\upsilon}{2}\left \| \textbf g - \psi(\textbf y^{(k+1)})\right \|_F^2, 
\end{aligned}
\end{equation}
\begin{equation}
\begin{aligned}
\textbf I^{(k+1)}=argmin_{\textbf I}{1}{2}\left \| \textbf I - \textbf I^{(k)}-\textbf g^{(k+1)}\right \|_F^2 +\xi f(\textbf I)), 
\end{aligned}
\end{equation}
where $\xi=\xi/\lambda$. Regarding as  the sub-problem of $\textbf y$ , it is solved by derivative descent method and then we have
\begin{equation}
\begin{aligned}
\textbf p^{(k+1)}=(\textbf I^{(L)}-\textbf {HI}^{(k)}+\upsilon \textbf H\textbf g^{(k)})/(1+\upsilon)
\end{aligned}
\end{equation}
To keep consistent with the original measurement, we assume that an initial condition $\textbf{Hg}^{(0)}=\textbf I^{L}$ is satisfied. In other words, the above formula is valid for $k=0$ with the condition $\textbf{Hg}^{(0)}=\textbf I^{L}$.

Regarding the sub-problem for g, the solution can be directly obtained:
\begin{equation}
\begin{aligned}
\textbf g^{(k+1)}=\frac{\upsilon \psi_(\textbf y^{(k+1)})}{\upsilon + \mu}
\end{aligned}
\end{equation}
Regarding as the regularization prior term, different selection of regularization priors result in different recovery results.  The regularization prior has an important effect on the final reconstruction. Among many priors for image reconstruction, including dictionary learning\cite{RN45}, low-rank\cite{RN27}, sparsity\cite{RN113}, and others\cite{RN114}, we use a simple TV-type regularizer to encourage the sparsity as an example:
\begin{equation}
\begin{aligned}
f(\textbf I)=\sum\nolimits_{j_2=2}^{J_2}\sum\nolimits_{j_1=2}^{J_1}\left|\textbf I(j_1,j_2)-\textbf f(j_1-1,j_2)\right|+{}\\ \left|\textbf f(j_1,j_2)-\textbf f(j_1,j_2-1)\right|), j_1=1,...,J_1;j_2=1,...J_2,
\end{aligned}
\end{equation}
where $J_1$ and $J_2$ represent the width and height of a reconstructed image, and the gradients on the image border are set to zero. Thus, $\textbf I^{(k+1)}$ can be updated as follows:
\begin{equation}
\begin{aligned}
\textbf I^{(k+1)}=argmin_{\textbf I}(\frac{1}{2})\left \| \textbf I - \textbf I^{(k)}-\textbf g^{(k+1)}\right \|_F^2{}\\ + \xi \sum_{j_2=2}^{J_2}\sum_{j_1=2}^{J_1}\left|\textbf I(j_1,j_2)-\textbf I(j_1-1,j_2)\right|+{}\\ \left|\textbf I(j_1,j_2)-\textbf I(j_1,j_2-1)\right|)
\end{aligned}
\end{equation}
Replacing $\textbf I(j_1,j_2 )-\textbf I(j_1-1,j_2)$ and $\textbf I(j_1,j_2 )-\textbf I(j_1,j_2-1)$ with $\textbf d_1 (j_1,j_2)$ and $\textbf d_2 (j_1,j_2)$ respectively, we have the unconstrained problem:
\begin{equation}
\begin{aligned}
\left \{\textbf I^{(k+1)}, \textbf d_1^{(k+1)}, \textbf d_2^{(k+1)}\right \}{}\\=argmin_{\left \{\textbf I, \textbf d_1, \textbf d_2\right\}}(\frac{1}{2})\left \| \textbf I - \textbf I^{(k)}-\textbf g^{(k+1)}\right \|_F^2+{}\\ \xi \sum_{j_2=2}^{J_2}\sum_{j_1=2}^{J_1}\left|\textbf d_1(j_1,j_2)-\textbf d_2(j_1,j_2)\right|+{}\\ \rho \sum_{j_2=2}^{J_2}\sum_{j_1=2}^{J_1}(\left| \textbf d_1(j_1,j_2)-(\textbf I(j_1,j_2)-\textbf I(j_1-1,j_2))\right|+{}\\ \left| \textbf d_2(j_1,j_2)-(\textbf I(j_1,j_2)-\textbf I(j_1,j_2-1))\right|)
\end{aligned}
\end{equation}
The above optimization problem can be solved by alternately minimizing the objective function. An FFT-based algorithm, FTVd\cite{Wang2007AFA}, is employed to find the solution. Note that there are two parameters in the above problem: $\rho$ and $\xi$. These parameters are made the same in this study, we can use the same variable $\delta$ to replace $\rho$ and $\xi$.

\begin{figure*}
	\centering
	\includegraphics[width = 1.0\textwidth]{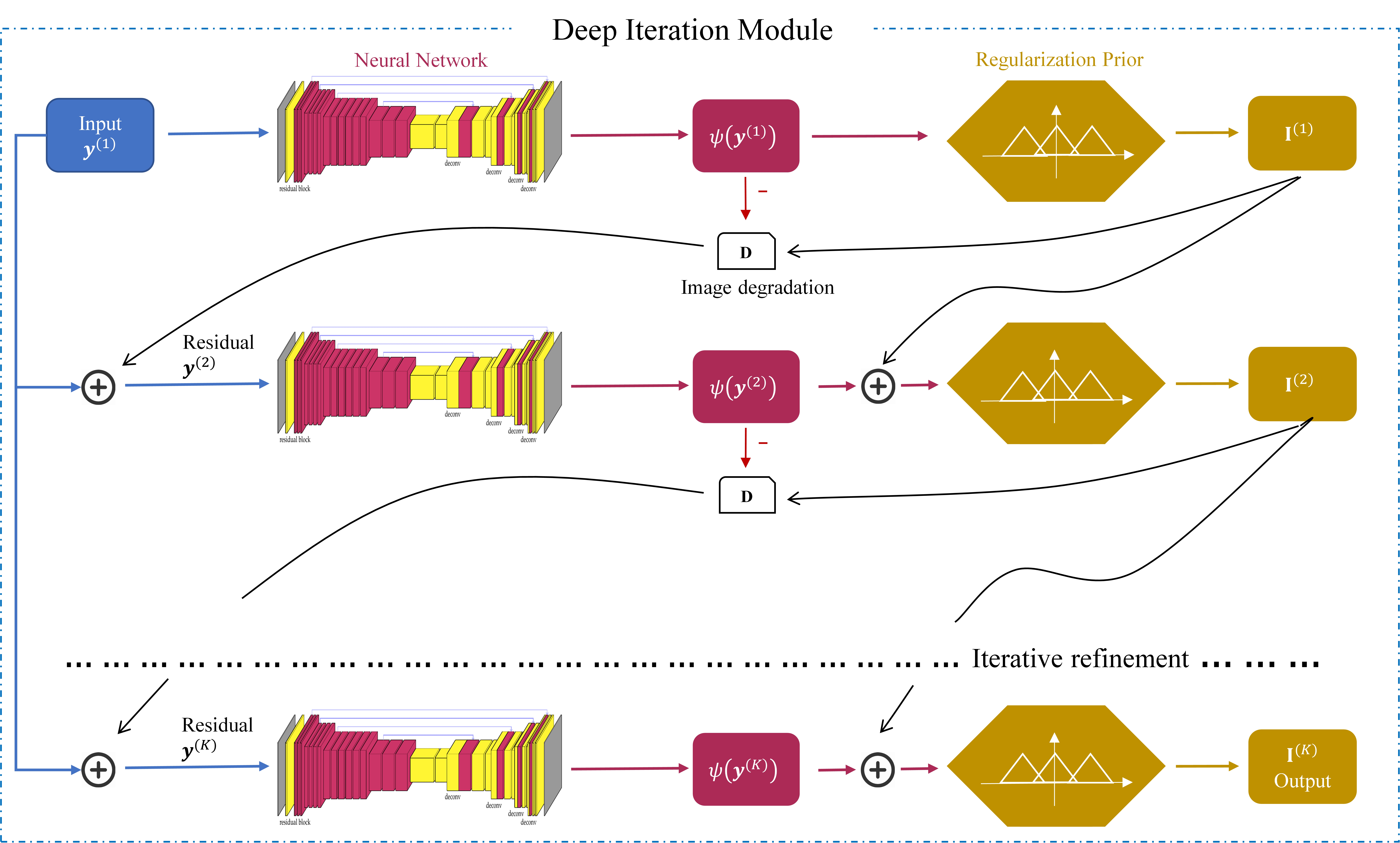}
	\caption{Architecture of deep iteration module. This module consists of four components: deep reconstruction, compressed sensing, image degradation mapping, and iterative refinement. $\textbf p^{(1)}$ is the original tomographic dataset, and $\textbf p^{(k)}, k=2, 3, …, K$, represents an estimated residual dataset in the $k^{th}$ iteration between $\textbf p^{(1)}$ and the currently reconstructed counterpart. $\Phi_\textbf w(\textbf p^{(k)})$ is an output of the deep reconstruction module, and $\textbf f(k)$ represents a reconstruction regularized via compressed sensing.}
	\label{Figure:3}
\end{figure*}

\begin{figure*}[htb]
	\centering
	\includegraphics[width = 0.9\textwidth]{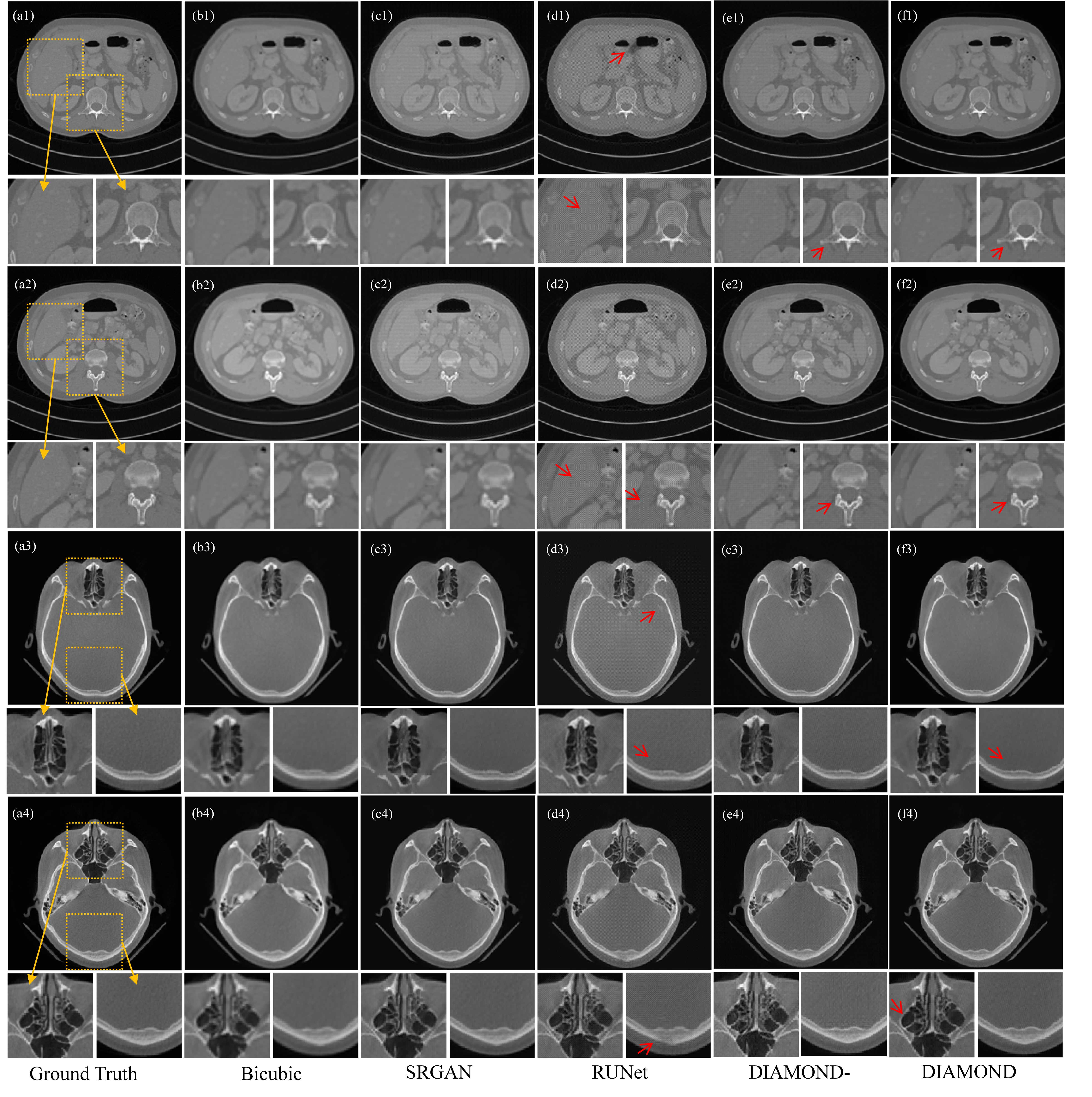}
	\caption{CT Image Super-resolution Results. The 1st-4th rows are random abdominal CT  slices and their corresponding ROIs from AAPM dataset. The5th-8th rows are random oral CT  slices and their corresponding ROIs from local hospital.}
	\label{Figure:2}
\end{figure*}

\begin{figure*}[htb]
	\centering
	\includegraphics[width = 0.9\textwidth]{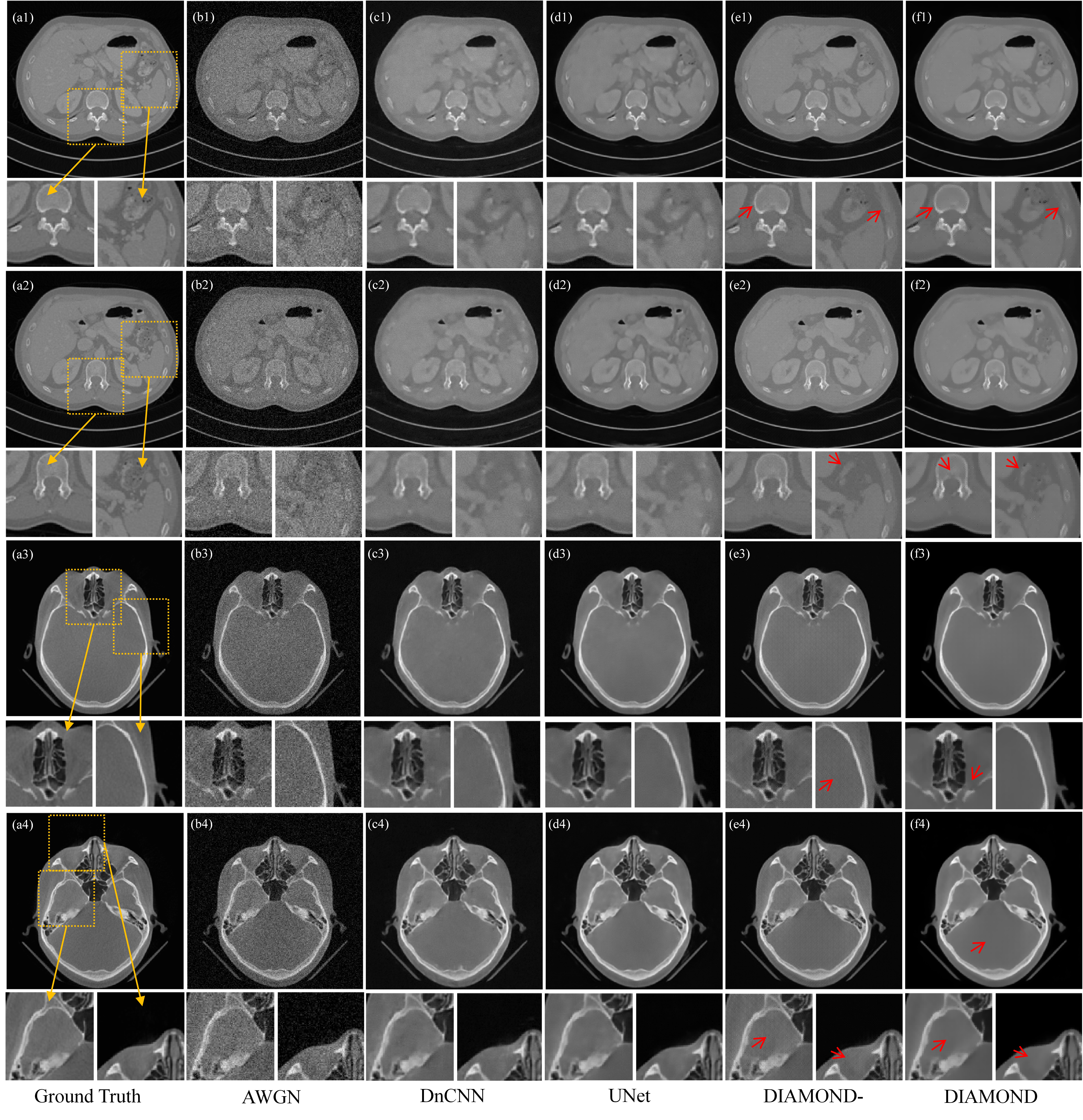}
	\caption{CT Image-denosing Results. The 1st-4th rows are random abdominal CT  slices and their corresponding ROIs from AAPM dataset. The 5th-8th rows are random oral CT  slices and their corresponding ROIs from Jiangsu Province Hospital, China.}
	\label{Figure:2}
\end{figure*}

\begin{figure}[htb]
	\centering
	\includegraphics[width = 0.45\textwidth]{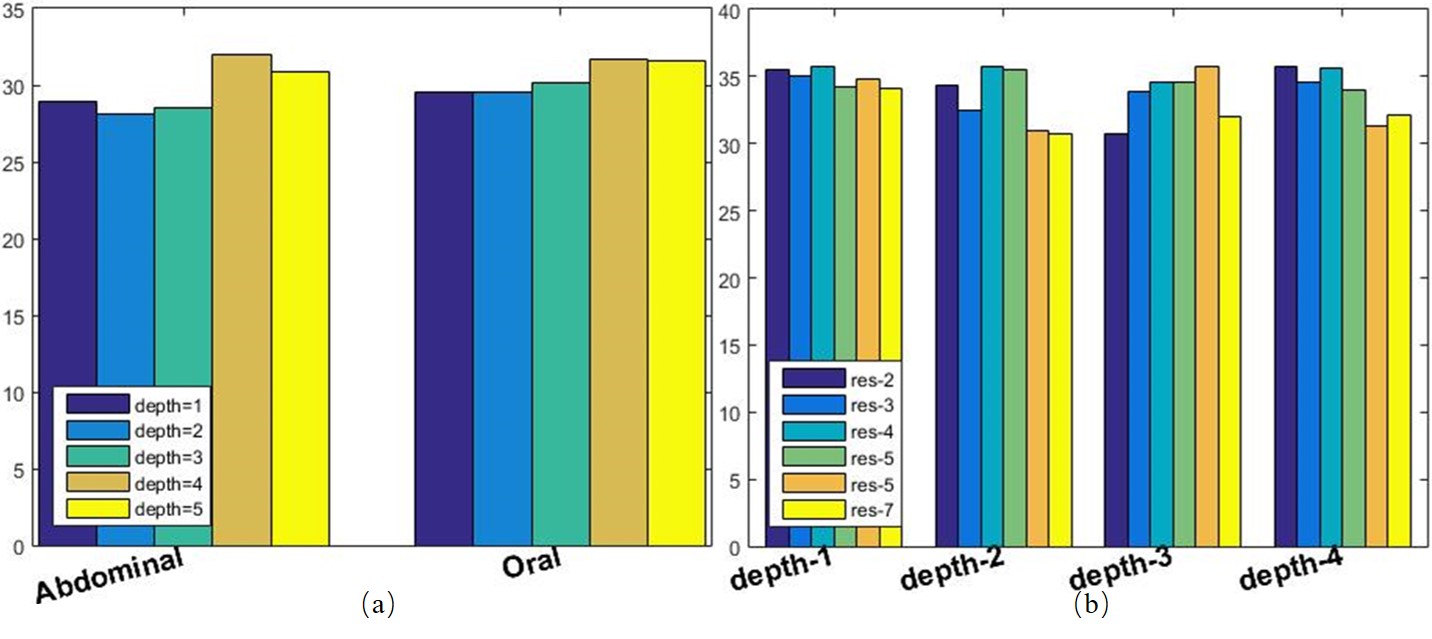}
	\caption{Histograms of ablation results on extracting depth and number of residual blocks. (a) shows PSNR values on two datasets considering different contracting depths. After confirming contracting depth(=4), (b) shows PSNR values on abdominal dataset considering different residual blocks}
	\label{Figure:5}
\end{figure}

\begin{figure*}[htb]
	\centering
	\includegraphics[width = 0.9\textwidth]{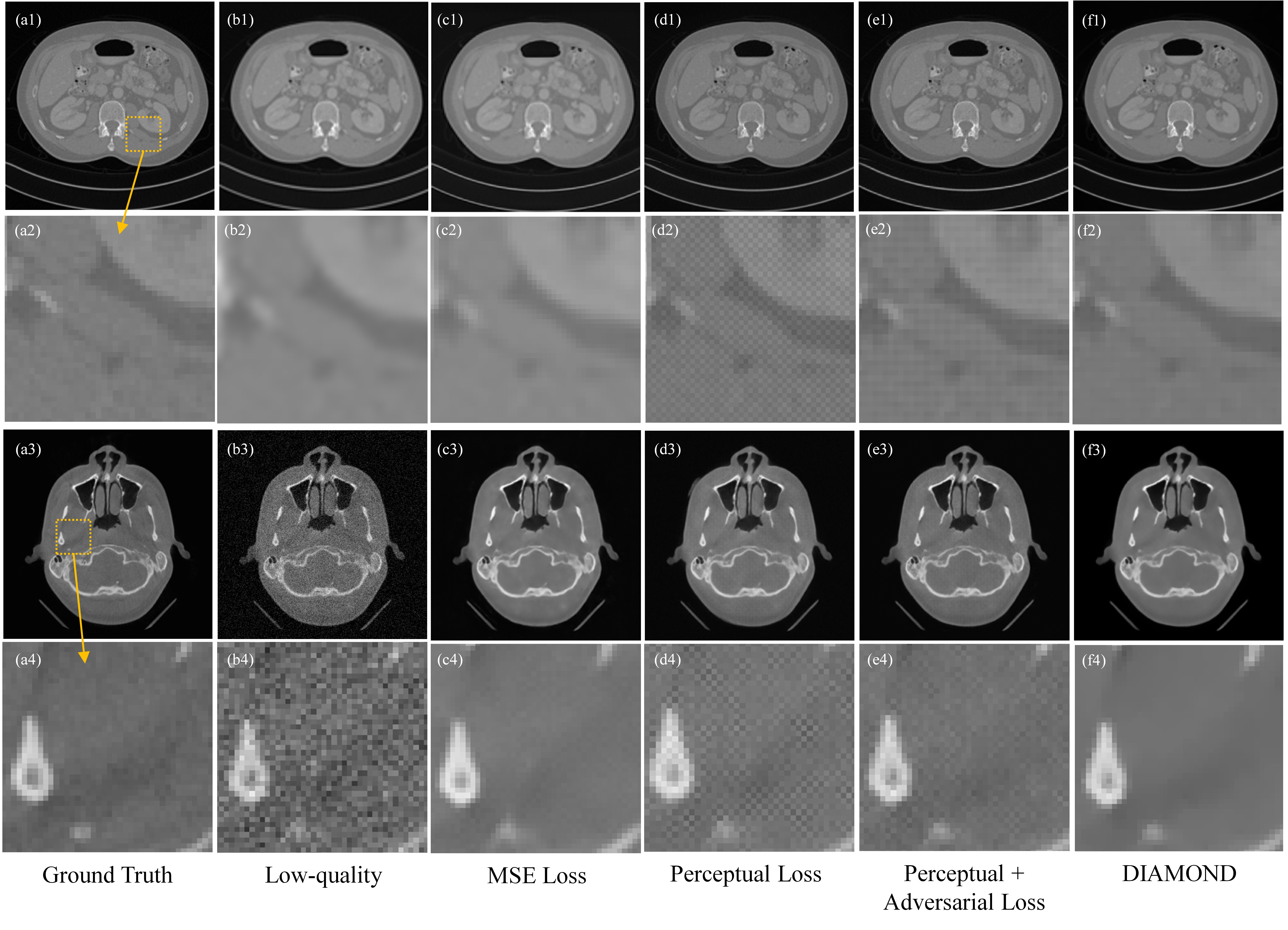}
	\caption{Ablation experiment results in loss functions. The 1st and 2nd rows are random abdominal CT slices for super-resolution tasks and their corresponding ROIs. The 3rd and 4th rows are are random oral CT slices for de-noising tasks and their corresponding ROIs.}
	\label{Figure:8}
\end{figure*}

\begin{figure*}[htb]
	\centering
	\includegraphics[width = 0.9\textwidth]{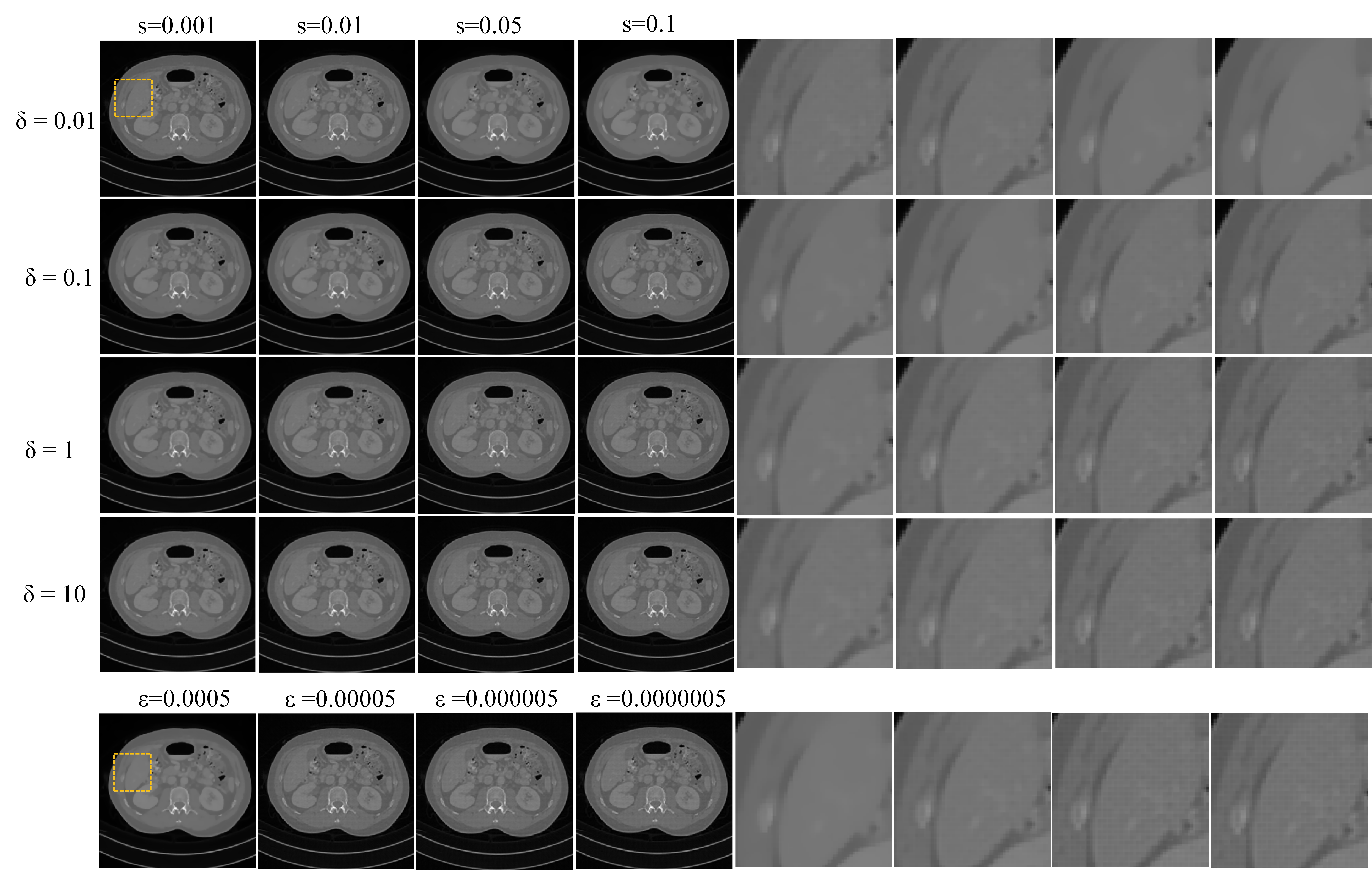}
	\caption{Super-resolution results (DIAMOND) with different parameters.ROIs are listed at the right together.After optimizing step s and $\delta$, the TV parameter $\varepsilon$ is optimized on the last row.}
	\label{Figure:6}
\end{figure*}

\begin{figure}[htb]
	\centering
	\includegraphics[width = 0.46\textwidth]{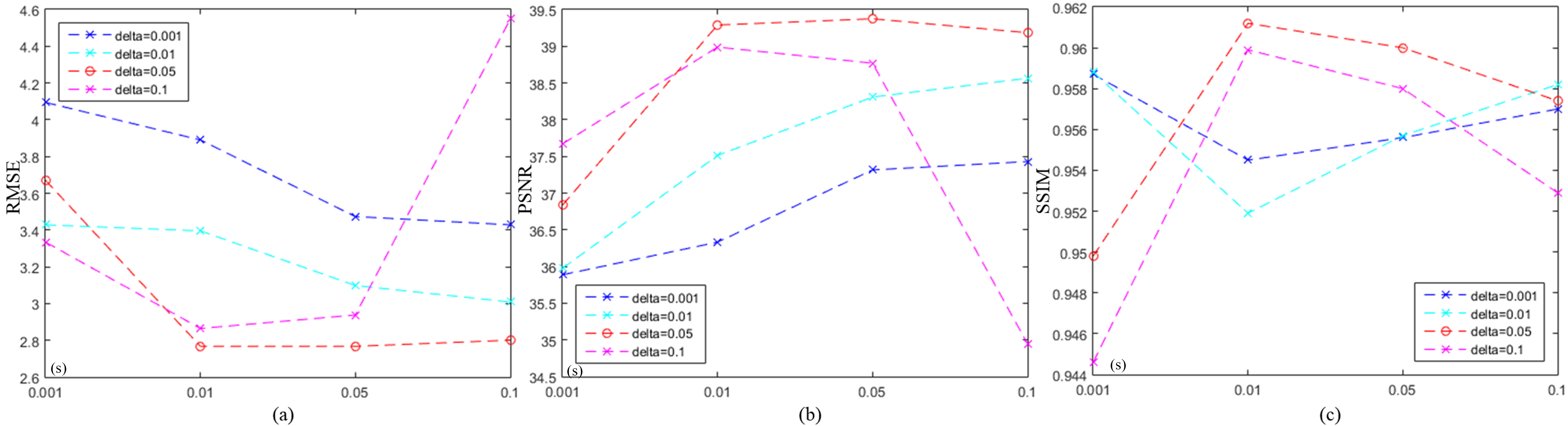}
	\caption{RMSE, PSNR, and SSIM line diagrams of Fig. 7 results(row 1-4).}
	\label{Figure:7}
\end{figure}
\begin{figure}[htb]
	\centering
	\includegraphics[width = 0.46\textwidth]{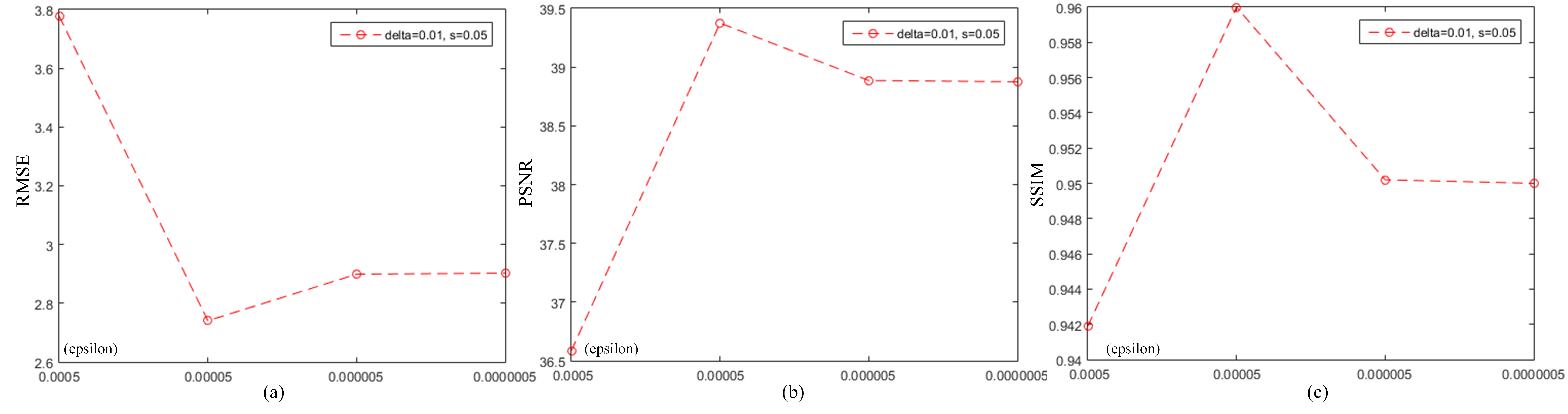}
	\caption{RMSE, PSNR, and SSIM line diagrams of Fig. 7 results(row 5).}
	\label{Figure:8}
\end{figure}

\textbf {Deep Iteration Module Mechanism}: As demonstrated in Fig. 2, the mechanism of the deep iteration module is based on the iterative refinement. The error feedback is essential to recover structural subtleties that can be lost using a single neural network. This mechanism helps effectively suppress mismatches and/or inconsistencies caused by existing deep learning methods \cite{RN32}\cite{RN33}.
The output of the neural network is combined with the data as the input to the DL reconstruction network. The trained neural network is employed to perform image recovery again so that one can obtain residual image and then add it to the previous recovery result. The deep learning network and compressed sensing at each iteration produce residual image for a gradually improved image recovery. It is easy to understand that the DL network is trained on original images but it may not directly produce an idea clean image that is consistent with the sparsity requirement by compressed sensing. This issue can be addressed with regularization prior in terms of total variation \cite{RN17}, low-rank \cite{RN18}, dictionary learning \cite{RN19}, etc. In this study, the anisotropic TV is employed to perform such task\cite{RN34}. 

\section{Experiments}
We implemented the proposed models using the Tensorflow framework. We use the python implementation of Bicubic interpolation, SRGAN, and RU-Net to do super-resolution tasks for fair comparisons. We also use the Tensorflow implementation of U-Net, DnCNN, and GAN to do denoising tasks. The performance of the proposed DIAMOND is evaluated on simulated and real datasets. For image super-resolution tasks, we first conduct simulated experiments to verify DIAMOND's mechanism in image super-resolution tasks. We use real datasets to furtherly prove the method's effectiveness. For image denoising tasks, we also follow the previous experimental process. All the experiments are implemented on Ubuntu (14 CPUs Intel Xeon E5-2683 v3, @2GHz, Titan X GPU, 12.0 GB VRAM, 64.0 GB RAM).

\textbf{Evaluation measures.} Four quantitative picture quality indices (PQI) are employed for performance evaluation, including root mean square error (RMSE), peak signal-to-noise ratio (PSNR), structure similarity index (SSIM). A smaller MSE value refers to a subtler deviation between the reconstructed image and reference image. A larger PSNR value means a higher image quality. A larger SSIM value reflects a higher similarity in image structures.

\textbf{Implementation details.} We implement and train our network using the Tensorflow framework. We use Adam optimizer to train the network for 200 epochs, 1100 iterations. The learning rate $l$ is halved every 100 epochs. The batch size $b$ is set according to the number of training data. More implementation details are listed in Table. 1.

\begin{table*}[htbp]
  \centering
    \begin{tabular}{ccccccccc}
    \multicolumn{8}{l}{Table 1. Parameter values for all experiments} &  \\
    \midrule
    \midrule
    \multirow{2}[2]{*}{} & \multicolumn{5}{c}{\multirow{2}[2]{*}{Deep Network Module}} & \multicolumn{3}{c}{\multirow{2}[2]{*}{Deep Iteration module}} \\
          & \multicolumn{5}{c}{}                  & \multicolumn{3}{c}{} \\
    \midrule
          &       & $\lambda$ & $\Lambda$ & $l$     & $b$     & $s$     & $\delta$ & $\varepsilon$ \\
    \multirow{2}[1]{*}{Denoising} & abdominal & 0.005 & 10    & 0.00005 & 16    & 0.0005 & -     & 0.0009 \\
          & oral  & 0.001 & 10    & 0.0001 & 48    & 0.0001 & -     & 0.0009 \\
    \midrule
    \multirow{2}[2]{*}{Super resolution} & abdominal & 0.005 & 10    & 0.00002 & 16    & 0.05  & 0.01  & 0.00005 \\
          & oral  & 0.001 & 10    & 0.00002 & 48    & 0.01 & 1 & 0.00025 \\
    \bottomrule
    \bottomrule
    \end{tabular}%
  \label{tab:addlabel}%
\end{table*}%

\subsection{Image Super-resolution Results}
In this study, an abdominal cavity CT dataset from AAPM competition is first used to compare all reconstruction methods' performance. After proving DIAMOND outperforms all other methods, we further apply our method to real oral cavity CT data from Jiangsu Province Hospital, China. The size of the original data is $256\times256$. We first downsample this image to $128\times128$, then do bicubic interpolation to recover its original size. Then we use the interpolated image as the input of the network module of DIAMOND. Mean square error(MSE), peak signal-to-noise ratio(PSNR), and structural similarity(SSIM) are employed to access the reconstruction results quantitatively. To reach an optimal performance of the proposed method, we modify hyper-parameter values empirically. For all methods, hyper-parameters' optimized values to minimize RMSE (also maximizing PSNR) have been selected. We now list all of them below in Table. 2.

To validate the performance of the proposed DIAMOND method for image super-resolution reconstruction, Fig. 3 shows the reconstruction results using all super-resolution methods. The downsampled scale is set as 2.To fairly compare the performance of all methods, the parameters have been optimized to obtain the best results. Results in the 5th column are obtained by the mere network module of DIAMOND and are named DIAMOND-. Pre-upsampling images are obtained from bicubic interpolation of downsampled counterparts. Fig. 3 demonstrates that the proposed method leads to images with better edge preservation and adequate feature discovery than those obtained with other methods. More specifically, pre-upsampling images suffer from severe blur and detail missing, as shown in Fig. 3(b1)-(b4). Circular artifacts are observed in RUNet results, as illustrated by Fig. 3(d1)-(d4). DIAMOND- achieves better results than the above methods in suppressing circular artifacts and restoring image details, which can be observed from extracted regions-of-interest (ROIs) in Fig. 3(e1)-(e4). Compared with DIAMOND-, the proposed method has better performance in subtle detail preservation, as pointed by arrows in ROIs in Fig. 3(f1)-(f4).

\newcommand{\tabincell}[2]{\begin{tabular}{@{}#1@{}}#2\end{tabular}}

\begin{table}[htbp]
  \centering
 \resizebox{0.45\textwidth}{!}{
    \begin{tabular}{ccccccc}
    \multicolumn{7}{l}{Table. 2 Results of different Super Resolution methods on two datasets} \\
    \midrule
    \midrule
    \multirow{5}[4]{*}{\tabincell{c}{AAPM\\abdominal\\CT data}} & \multirow{2}[2]{*}{} & \multirow{2}[2]{*}{Bicubic} & \multirow{2}[2]{*}{SRGAN} & \multirow{2}[2]{*}{RUNet} & \multirow{2}[2]{*}{DIAMOND-} & \multirow{2}[2]{*}{DIAMOND} \\
          &       &       &       &       &       &  \\
\cmidrule{2-7}          & RMSE  & 6.9677 & 8.0404 & 7.9280  & 4.2147 & \textbf{2.7403} \\
          & PSNR  & 31.2399 & 30.0269 & 29.7839 & 35.6635 & \textbf{39.3748} \\
          & SSIM  & 0.8983 & 0.9264 & 0.7342 & 0.9497 & \textbf{0.9600} \\
    \midrule
    \multirow{5}[4]{*}{\tabincell{c}{real oral\\CT data}} & \multirow{2}[2]{*}{} & \multirow{2}[2]{*}{Bicubic} & \multirow{2}[2]{*}{SRGAN} & \multirow{2}[2]{*}{RUNet} & \multirow{2}[2]{*}{DIAMOND-} & \multirow{2}[2]{*}{DIAMOND} \\
          &       &       &       &       &       &  \\
\cmidrule{2-7}          & RMSE  & 11.2241  & 7.1176  & 5.3530  & 4.6239  & \textbf{4.4294}  \\
          & PSNR  & 27.0790  & 31.0842  & 33.2968  & 35.2895  & \textbf{35.6923}  \\
          & SSIM  & 0.7856  & 0.8503  & 0.8406  & 0.9037  & \textbf{0.9186}  \\
    \bottomrule
    \bottomrule
    \end{tabular}}
  \label{tab:addlabel}%
\end{table}%

\begin{table}[htbp]
  \centering
  \resizebox{0.45\textwidth}{!}{
    \begin{tabular}{ccccccc}
    \multicolumn{7}{l}{Table. 3 Results of different Denoising methods on two datasets} \\
    \midrule
    \midrule
    \multirow{5}[4]{*}{\tabincell{c}{AAPM\\abdominal\\CT data}} & \multirow{2}[2]{*}{} & \multirow{2}[2]{*}{AWGN} & \multirow{2}[2]{*}{DnCNN} & \multirow{2}[2]{*}{Unet} & \multirow{2}[2]{*}{DIAMOND-} & \multirow{2}[2]{*}{DIAMOND} \\
          &       &       &       &       &       &  \\
\cmidrule{2-7}          & RMSE  & 13.6934  & 5.6987  & 5.1152  & 5.1148  & \textbf{4.6276}  \\
          & PSNR  & 25.4007  & 33.0254  & 33.9536  & 33.9542  & \textbf{34.8236}  \\
          & SSIM  & 0.4124  & 0.7601  & 0.8638  & 0.8644  & \textbf{0.8939}  \\
    \midrule
    \multirow{5}[4]{*}{\tabincell{c}{real oral\\CT data}} & \multirow{2}[2]{*}{} & \multirow{2}[2]{*}{AWGN} & \multirow{2}[2]{*}{DnCNN} & \multirow{2}[2]{*}{UNet} & \multirow{2}[2]{*}{DIAMOND-} & \multirow{2}[2]{*}{DIAMOND} \\
          &       &       &       &       &       &  \\
\cmidrule{2-7}          & RMSE  & 13.4655  & 5.0630  & 5.4125  & 4.3722  & \textbf{3.7084}  \\
          & PSNR  & 25.5467  & 34.0429  & 33.6077  & 35.3486  & \textbf{36.7558}  \\
          & SSIM  & 0.4316  & 0.9040  & 0.9296  & 0.9277  & \textbf{0.9584}  \\
    \bottomrule
    \bottomrule
    \end{tabular}}
  \label{tab:addlabel}%
\end{table}%

Table. 2 shows the quantitative results (RMSE, PSNR, SSIM) concerning super-resolution reconstructions in Fig. 2. In Fig. 3, we show two slices for both abdominal data and oral data. In Table. 2, the quantitative results of the two slices are averaged. It can be figured out that our proposed method has the smallest RMSE value and highest PSNR and SSIM value, meaning that our proposed method can achieve the nearest distance from ground truth, suppressing noise and preserving subtle details. It should be mentioned that SRGAN and RUNet results are not quantitatively better than bicubic interpolation results though they can maintain image structures and optimize fine details much better visually. That is mainly due to the artifacts introduced by perceptual loss functions. In our method, we manage to remove artifact pollution, which results in a quantitative promotion.

\subsection{Image Denoising Results}
To validate the performance of the proposed DIAMOND method for image denoising tasks, we need to prepare a training dataset of input-output pairs $\{(y_i,M_i;x_i)\}_{i=1}^N$. Here, $y_i$ is obtained by adding Additive White Gaussian Noise (AWGN) to latent image $x_i$ and $M_i$ is the noise level map. The reason to use AWGN to generate the training dataset is two-fold. First, AWGN is a natural choice when there is no specific prior information on the noise source. Second, real-world noise can be approximated as locally AWGN. We found that the learned model still works on real noisy images.

We compare our DIAMOND method with DnCNN, U-Net, and DIAMOND- methods on the same datasets. All methods are trained through residual learning. The RMSE, PSNR, and SSIM values are shown in Table. 3, which undoubtedly indicates that our method outperforms others. The results are visualized in Fig. 4, showing that the DIAMOND method can effectively remove AWGN without generating annular artifacts. Fig. 4(d1)-(d4) shows that U-Net results destroy image details though removing noise. Fig. 4(e1)-(e4) shows that DIAMOND- do well in removing AWGN and preserve image structures. However, this method fails to suppress annular artifacts, which degrades image quality. Equipped with a post-processing deep iteration module and a GAN network, the proposed method have an obvious advantage in removing noise, suppressing artifacts, and preserving delicate features, as is shown in Fig. 4 (f1)-(f4). Moreover, it can be readily illustrated in Table. 3 that the proposed method outperform all others in all three indexes.

Fig. 5 analyzes the convergence speed of the proposed method in Super-resolution tasks. It can be figured out that our method converges at tens of steps in different datasets. Also, their index values (RMSE, PSNR, and SSIM) are promoted during the process. Fig. 6 shows the convergence curves in denoising tasks. The deep iteration module performs as a useful tool to promote the recovery images, which reduces RMSE value and increases PSNR and SSIM values during iterations.

\subsection{Ablation Experiments}
This section compares the contracting/expanding depth of the generator network, the number of residual blocks in the contracting/expanding path, and the loss functions, respectively. First, considering the generator architecture in Fig. 1(a), we analyze how its depth influences output image quality. Since the input patch is set as $64\times64$, the contracting depth is possible to be from one to five and we compare the results in denoising tasks using the generator network. Second, after confirming the optimal contracting depth, the number of optimal residual blocks for each depth is further discussed. Third, pixel-wise loss functions, perceptual loss functions, and adversarial loss functions are compared in the area of feature preservation and artifact introduction.
\subsubsection{Depth of Contracting/Expanding Path}
We modify the generator network and achieve several results to analyze how contracting/expanding depth affects image de-noising results. Table. 4 shows the PSNR value of de-noising images regarding contracting/expanding depth. It can be figured out that both datasets can reach the highest PSNR values when the depth is set as 4. The histogram in Fig. 5 (a) also proves this discussion.
\subsubsection{Number of Residual Blocks}
On the basis that the optimal contracting depth is four, further ablation experiments are performed to analyze each depth's optimal number of residual blocks in super-resolution tasks. It should be pointed out that at least two residual blocks should exist on each path, both to extract and transmit features, meanwhile the last one to increase the number of feature maps. Moreover, when the number of residual blocks on a certain depth is discussed, all the other depths achieve the residual blocks' optimal number.

Table 5 and Fig. 5 (b) show the experimental PSNR results on each depth's number of residual blocks. To reach the highest PSNR value, four residual blocks are in the first depth, four residual blocks in the second depth, six residual blocks in the third depth, and two residual blocks in the fourth depth.

\begin{table}[htbp]
  \centering
  \resizebox{0.45\textwidth}{!}{
    \begin{tabular}{rrrrrr}
    \multicolumn{6}{l}{Table 4. Analysis on depth of contracting/ecpanding path (/dB)} \\
    \midrule
    \midrule
    \multicolumn{1}{l}{\multirow{2}[2]{*}{Depth}} & \multicolumn{1}{c}{\multirow{2}[2]{*}{1}} & \multicolumn{1}{c}{\multirow{2}[2]{*}{2}} & \multicolumn{1}{c}{\multirow{2}[2]{*}{3}} & \multicolumn{1}{c}{\multirow{2}[2]{*}{4}} & \multicolumn{1}{c}{\multirow{2}[2]{*}{5}} \\
          &       &       &       &       &  \\
    \midrule
    \multicolumn{1}{c}{\multirow{2}[1]{*}{Abdominal}} & \multicolumn{1}{c}{\multirow{2}[1]{*}{28.8802}} & \multicolumn{1}{c}{\multirow{2}[1]{*}{28.1417}} & \multicolumn{1}{c}{\multirow{2}[1]{*}{28.5645}} & \multicolumn{1}{c}{\multirow{2}[1]{*}{\textbf{32.0444}}} & \multicolumn{1}{c}{\multirow{2}[1]{*}{30.8867}} \\
          &       &       &       &       &  \\
    \multicolumn{1}{c}{\multirow{2}[1]{*}{Oral}} & \multicolumn{1}{c}{\multirow{2}[1]{*}{29.5665}} & \multicolumn{1}{c}{\multirow{2}[1]{*}{29.5717}} & \multicolumn{1}{c}{\multirow{2}[1]{*}{30.1515}} & \multicolumn{1}{c}{\multirow{2}[1]{*}{\textbf{31.6979}}} & \multicolumn{1}{c}{\multirow{2}[1]{*}{31.6150}} \\
          &       &       &       &       &  \\
    \midrule
          &       &       &       &       &  \\
    \end{tabular}}
  \label{tab:addlabel}%
\end{table}%
\begin{table}[htbp]
  \centering
  \resizebox{0.45\textwidth}{!}{
    \begin{tabular}{ccccccc}
    \multicolumn{7}{l}{Table 5. Analysis on number of residual blocks (/dB)} \\
    \midrule
    \midrule
    \multicolumn{1}{l}{\multirow{2}[2]{*}{       }} & \multirow{2}[2]{*}{res-2} & \multirow{2}[2]{*}{res-3} & \multirow{2}[2]{*}{res-4} & \multirow{2}[2]{*}{res-5} & \multirow{2}[2]{*}{res-6} & \multirow{2}[2]{*}{res-7} \\
          &       &       &       &       &       &  \\
    \midrule
    \multirow{2}[1]{*}{depth-1} & \multirow{2}[1]{*}{35.5138 } & \multirow{2}[1]{*}{35.0753 } & \multirow{2}[1]{*}{\textbf{35.7260} } & \multirow{2}[1]{*}{34.2440 } & \multirow{2}[1]{*}{34.7818 } & \multirow{2}[1]{*}{34.1120 } \\
          &       &       &       &       &       &  \\
    \multirow{2}[0]{*}{depth-2} & \multirow{2}[0]{*}{34.3963 } & \multirow{2}[0]{*}{32.4851 } & \multirow{2}[0]{*}{\textbf{35.7260} } & \multirow{2}[0]{*}{35.5175 } & \multirow{2}[0]{*}{30.9810 } & \multirow{2}[0]{*}{30.7861 } \\
          &       &       &       &       &       &  \\
    \multirow{2}[0]{*}{depth-3} & \multirow{2}[0]{*}{30.7797 } & \multirow{2}[0]{*}{33.8599 } & \multirow{2}[0]{*}{34.2451 } & \multirow{2}[0]{*}{34.5942 } & \multirow{2}[0]{*}{\textbf{35.7260} } & \multirow{2}[0]{*}{32.0646 } \\
          &       &       &       &       &       &  \\
    \multirow{2}[1]{*}{depth-4} & \multirow{2}[1]{*}{\textbf{35.7260} } & \multirow{2}[1]{*}{34.5877 } & \multirow{2}[1]{*}{35.6352 } & \multirow{2}[1]{*}{33.9648 } & \multirow{2}[1]{*}{31.2647 } & \multirow{2}[1]{*}{32.1052 } \\
          &       &       &       &       &       &  \\
    \bottomrule
    \end{tabular}}
  \label{tab:addlabel}%
\end{table}%

\subsubsection{Loss Functions}
In this section, three loss functions are discussed: pixel-wise loss, perceptual loss, and adversarial loss.

The definition of pixel-wise loss is in Eq. (3). This loss function calculates the pixel loss between the predicted images and the target images. Standard pixel-wise loss functions, such as MSE or L2 loss, can be mostly applied between each pair of predicted and the target pixels. Since these loss functions evaluate each pixel vector separately and then average all pixels, they assert that the same learning is done for each pixel in the image. Pixel-wise loss is widely used in image recovery tasks. However, pixel-wise loss concentrates on pixel-level similarity and sometimes misses the overall image effect or losses subtle image details, shown in Fig. 6 (b1)-(b4). (b1) is the super-resolution result of one abdominal CT slice contrained by MSE loss, (b2) is its corresponding ROI. In this project, MSE loss hardly works in super-resolution reconstruction tasks. (b3) is the de-noising result of one oral cavity CT slice constrained by MSE loss, (b2) is its corresponding ROI. MSE loss can remove noise and promote image qualities to some extent. However, compared with (d3) and (d4), it fails to preserve subtle details.

The perceptual loss is defined in Eq. (4) and (5). It compares two different images that look similar, such as the same image at different resolutions. Even though the images are very similar in these cases, the pixel-level loss function will output a considerable error value. However, the perceptual loss function compares high-level perception and semantic differences between images and is good at preserving image details and delicate structures in image super-resolution tasks.
Nevertheless, deep networks constrained by perceptual loss tends to introduce artifacts into reconstructions. From Fig. 6 (d1)-(d4), we can see that though minimizing perceptual loss is beneficial to recover image details, it introduces annular artifacts into reconstructed images and degrades visual effects.

To suppress artifacts introduced by the perceptual loss, the adversarial loss is introduced into the overall loss functions with a weight parameter $\lambda$ (see Eq. (2) and Eq. (6)). The discriminator can be valid to capture the potential attributes of high-resolution images. Compared with Fig. 6 (d2) and (d4), artifacts in (e2) and (e4) are visibly reduced.

Finally, It should be mentioned that some artifacts remain in Fig. 6 (e2) and (e4). Our Proposed method further removes artifacts and predicts images with the best visual effect(see Fig. 6 (f1)-(f4)).

\subsubsection{Deep Iteration Module Parameters}
In this section, we provide some suggestions on parameter selections of the deep iteration module. There are three parameters in all in this module: ADMM optimization parameter $\delta$, iterative step $s$ and TV parameter $\varepsilon$. The results are shown in Fig. 7 and Fig. 8.

With the decrase of $s$, iterative results tend to have more delicate results. However, the smaller step will not only increase artifacts but also slow down convergence. Increasing $\delta$ can subtly preserve image details. However, a more extensive $\delta$ will also bring artifacts. Increasing $\varepsilon$ can suppress artifacts and smooth the whole image structure. It is significant to tradeoff between these parameters while performing deep iteration operations.

\section{Conclusion}
In this study, we propose a novel strategy to solve general medical image restoration tasks. Our contributions are three folds: First, we put forward a novel GAN network with multi-level residual blocks and WGAN-GP training. Second, a deep iteration module combines deep learning with compressed sensing and promote restoration iteratively. Third, we incorporate the perceptual loss into the loss function and manage to suppress artifacts introduced by that loss function.

Medical imaging is a widely applied field, and a distinct medical image is helpful to medical diagnosis in many ways. However, medical images are sometimes polluted by noise or cannot reach resolution demands. An effective way to restore these polluted images and reach a satisfying image quality both in visual effect and indexes is highly significant in such cases. Our proposed method can restore the previous images and achieve reasonably outstanding performance in different datasets compared with competing methods. Moreover, we elaborately compare our method with the network part of our method to point out that the DIAMOND strategy performs better than the mere network. Also, the proposed network performs better than state-of-art methods, which is shown in Section 3.

It is also important to point out that our method has some shortcomings:

1. The proposed method only deals with $2\times$ super-resolution task and noise-level-15 de-noising task. The proposed method only deals with $2\times$ super-resolution task and noise-level-15 de-noising task. The perceptual loss function restricts its performance in more difficult recovery tasks. More loss function constraints, like $L_1$ loss functions, can be combined into current loss functions to preserve image structures.

2. This strategy consists of two steps, and for the iteration module, the computational cost is relatively high. Further research should focus on simpler regularization priors to speed up convergence.

3. Whether this strategy applies to other image restoration tasks is worthy of a try. Other image recovery tasks, such as image deblurring and image inpainting, can be taken into consideration.

In future research, we will conduct further experiments based on the three points above.


%

\ifCLASSOPTIONcaptionsoff
  \newpage
\fi



%

\bibliographystyle{IEEEtran}

\bibliography{reference.bib}

%

\end{document}